




\font\twelverm=cmr10 scaled 1200    \font\twelvei=cmmi10 scaled 1200
\font\twelvesy=cmsy10 scaled 1200   \font\twelveex=cmex10 scaled 1200
\font\twelvebf=cmbx10 scaled 1200   \font\twelvesl=cmsl10 scaled 1200
\font\twelvett=cmtt10 scaled 1200   \font\twelveit=cmti10 scaled 1200

\skewchar\twelvei='177   \skewchar\twelvesy='60


\def\twelvepoint{\normalbaselineskip=12.4pt
  \abovedisplayskip 12.4pt plus 3pt minus 9pt
  \belowdisplayskip 12.4pt plus 3pt minus 9pt
  \abovedisplayshortskip 0pt plus 3pt
  \belowdisplayshortskip 7.2pt plus 3pt minus 4pt
  \smallskipamount=3.6pt plus1.2pt minus1.2pt
  \medskipamount=7.2pt plus2.4pt minus2.4pt
  \bigskipamount=14.4pt plus4.8pt minus4.8pt
  \def\rm{\fam0\twelverm}          \def\it{\fam\itfam\twelveit}%
  \def\sl{\fam\slfam\twelvesl}     \def\bf{\fam\bffam\twelvebf}%
  \def\mit{\fam 1}                 \def\cal{\fam 2}%
  \def\tt{\twelvett}
  \textfont0=\twelverm   \scriptfont0=\tenrm   \scriptscriptfont0=\sevenrm
  \textfont1=\twelvei    \scriptfont1=\teni    \scriptscriptfont1=\seveni
  \textfont2=\twelvesy   \scriptfont2=\tensy   \scriptscriptfont2=\sevensy
  \textfont3=\twelveex \scriptfont3=\twelveex \scriptscriptfont3=\twelveex
  \textfont\itfam=\twelveit
  \textfont\slfam=\twelvesl
  \textfont\bffam=\twelvebf \scriptfont\bffam=\tenbf
  \scriptscriptfont\bffam=\sevenbf
  \normalbaselines\rm}



\def\beginlinemode{\endmode
  \begingroup\parskip=0pt \obeylines\def\\{\par}\def\endmode{\par\endgroup}}
\def\beginparmode{\endmode
  \begingroup \def\endmode{\par\endgroup}}
\let\endmode=\par
{\obeylines\gdef\
{}}
\def\singlespace{\baselineskip=\normalbaselineskip}

\def\oneandahalfspace{\baselineskip=\normalbaselineskip
  \multiply\baselineskip by 3 \divide\baselineskip by 2}
\def\doublespace{\baselineskip=\normalbaselineskip \multiply\baselineskip by 2}

\newcount\firstpageno
\firstpageno=2
\footline={\ifnum\pageno<\firstpageno{\hfil}\else{\hfil\twelverm\folio\hfil}
\fi}
\def\toppageno{\global\footline={\hfil}\global\headline
  ={\ifnum\pageno<\firstpageno{\hfil}\else{\hfil\twelverm\folio\hfil}\fi}}
\let\rawfootnote=\footnote              
\def\footnote#1#2{{\rm\singlespace\parindent=0pt\parskip=0pt
  \rawfootnote{#1}{#2\hfill\vrule height 0pt depth 6pt width 0pt}}}
\def\raggedcenter{\leftskip=4em plus 12em \rightskip=\leftskip
  \parindent=0pt \parfillskip=0pt \spaceskip=.3333em \xspaceskip=.5em
  \pretolerance=9999 \tolerance=9999
  \hyphenpenalty=9999 \exhyphenpenalty=9999 }
\def\dateline{\rightline{\ifcase\month\or
  January\or February\or March\or April\or May\or June\or
  July\or August\or September\or October\or November\or December\fi
  \space\number\year}}
\def\received{\vskip 3pt plus 0.2fill
 \centerline{\sl (Received\space\ifcase\month\or
  January\or February\or March\or April\or May\or June\or
  July\or August\or September\or October\or November\or December\fi
  \qquad, \number\year)}}


\hsize=6.5truein
\hoffset=0.05truein    
\vsize=8.9truein
\voffset=0.05truein   
\parskip=\medskipamount
\def\\{\cr}
\twelvepoint            
\doublespace            
\overfullrule=0pt       





\def\title                      
  {\null\vskip 3pt plus 0.1fill
   \beginlinemode \doublespace \raggedcenter \bf}

\def\author                     
  {\vskip 3pt plus 0.25fill \beginlinemode
   \singlespace \raggedcenter}

\def\affil                      
  {\vskip 3pt plus 0.1fill \beginlinemode
   \oneandahalfspace \raggedcenter \sl}

\def\abstract                   
  {\vskip 3pt plus 0.2fill \beginparmode
   \oneandahalfspace ABSTRACT: }

\def\resume                   
  {\vskip 3pt plus 0.2fill \beginparmode
   \oneandahalfspace RESUME: }

\def\endtopmatter               
  {\endpage                     
   \body}

\def\body                       
  {\beginparmode}               

\def\head#1{                    
  \goodbreak\vskip 0.5truein    
  {\immediate\write16{#1}
   \raggedcenter \uppercase{#1}\par}
   \nobreak\vskip 0.25truein\nobreak}

\def\beneathrel#1\under#2{\mathrel{\mathop{#2}\limits_{#1}}}

\def\refto#1{$^{#1}$}    

\def\references         
  {\head{References}      
   \beginparmode
   \frenchspacing \parindent=0pt \leftskip=1truecm
   \parskip=8pt plus 3pt \everypar{\hangindent=\parindent}}

\def\bibliog         
  {      
   \beginparmode
   \frenchspacing \parindent=0pt \leftskip=1truecm
   \parskip=8pt plus 3pt \everypar{\hangindent=\parindent}}

\gdef\refis#1{\item{#1.\ }}                     

\gdef\journal#1, #2, #3, 1#4#5#6{       
    {\sl #1~}{\bf #2}, #3 (1#4#5#6)}        

\def\prb{\journal Phys. Rev. B, }

\def\prl{\journal Phys. Rev. Lett., }

\def\jpc{\journal J. Phys. C, }

\def\jmmm{\journal J. Magn. Mag. Mat., }

\def\endreferences{\body}

\def\figurecaptions             
  {\endpage
   \beginparmode
   \head{Figure Captions}
}

\def\endfigurecaptions{\body}

\def\tablecaptions             
  {\endpage
   \beginparmode
   \head{Table Captions}
}

\def\endpage                    
  {\vfill\eject}

\def\endpaper                   
  {\endmode\vfill\supereject}

\def\endit
  {\endpaper\end}


\def\heading                            
  {\vskip 0.5truein plus 0.1truein      
   \beginparmode \def\\{\par} \parskip=0pt \singlespace \raggedcenter}

\def\subheading                         
  {\vskip 0.25truein plus 0.1truein     
   \beginlinemode \singlespace \parskip=0pt \def\\{\par}\raggedcenter}

\def\tag#1$${\eqno(#1)$$}

\def\align#1$${\eqalign{#1}$$}

\def\aligntag#1$${\gdef\tag##1\\{&(##1)\cr}\eqalignno{#1\\}$$
  \gdef\tag##1$${\eqno(##1)$$}}

\def\endaligntag{}

\def\overset#1\to#2{{\mathop{#2}^{#1}}}
\def\underset#1\to#2{{\mathop{#2}_{#1}}}


\def\ref#1{Ref.~#1}                     
\def\Ref#1{Ref.~#1}                     
\def\[#1]{[\cite{#1}]}
\def\cite#1{{#1}}
\def\(#1){(\call{#1})}
\def\call#1{{#1}}
\def\taghead#1{}
\def\frac#1#2{{#1 \over #2}}

\def\12{{1\over2}}

\def\sla{\raise.15ex\hbox{$/$}\kern-.57em}
\def\leaderfill{\leaders\hbox to 1em{\hss.\hss}\hfill}
\def\twiddle{\lower.9ex\rlap{$\kern-.1em\scriptstyle\sim$}}
\def\bigtwiddle{\lower1.ex\rlap{$\sim$}}
\def\gtwid{\mathrel{\raise.3ex\hbox{$>$\kern-.75em\lower1ex\hbox{$\sim$}}}}
\def\ltwid{\mathrel{\raise.3ex\hbox{$<$\kern-.75em\lower1ex\hbox{$\sim$}}}}
\def\square{\kern1pt\vbox{
\hrule height 1.2pt\hbox{\vrule width 1.2pt\hskip 3pt
\vbox{\vskip 6pt}\hskip 3pt\vrule width 0.6pt}\hrule height 0.6pt}\kern1pt}
\def\tdot#1{\mathord{\mathop{#1}\limits^{\kern2pt\ldots}}}

\def\pmb#1{\setbox0=\hbox{#1}%
  \kern-.025em\copy0\kern-\wd0
  \kern  .05em\copy0\kern-\wd0
  \kern-.025em\raise.0433em\box0 }

\def \ra{\rangle}

\def\dg{{^
{\dag}}}

\def\ra{\rangle}

\def\1{{\bf 1}}
\def\2{{\bf 2}}

\def\rarrow{\rightarrow}

\def\ul{\underline}

\def\vk{\vec k}
\def\iom{i \omega _n}
\def\ell{{\it l } {\rm n}}

\def\si{\sigma}

\def\cx2{\sqrt{c^2_x+c^2_y}}

\def\gkk{\gamma _{\vec k}}
\def\gk2{\gkk ^2}
\def\dw{\downarrow}
\def\up{\uparrow}
\def\gtappr{{{\lower4pt\hbox{$>$} } \atop \widetilde{ \ \ \ }}}
\def\ltappr{{{\lower4pt\hbox{$<$} } \atop \widetilde{ \ \ \ }}}

\def\pbar{{\partial\kern-1.2ex\raise0.25ex\hbox{/}}}

\def\up{\uparrow}
\def\dw{\downarrow}

\def\dg{{^{\dag}}}

\def\ra{\rangle}

\def\1{{\bf 1}}
\def\2{{\bf 2}}

\def\rarrow{\rightarrow}

\def\ul{\underline}

\def\vk{\vec k}
\def\vq{\vec q}
\def\iom{i \omega _n}
\def\ell{{\it l } {\rm n}}

\def\si{\sigma}

\def\cx2{\sqrt{c^2_x+c^2_y}}

\def\gkk{\gamma _{\vec k}}
\def\gk2{\gkk ^2}
\def\gtappr{{{\lower4pt\hbox{$>$} } \atop \widetilde{ \ \ \ }}}
\def\ltappr{{{\lower4pt\hbox{$<$} } \atop \widetilde{ \ \ \ }}}

\def\refto#1{$^{#1}$}

\def\eps{\epsilon}
\def\3he{{$^3${\rm He}}}


\def\slD{\raise.15ex\hbox{$/$}\kern-.57em\hbox{$D$}}
\def\dsl{\raise.15ex\hbox{$/$}\kern-.57em\hbox{$\Delta$}}
\def\slp{{\raise.15ex\hbox{$/$}\kern-.57em\hbox{$\partial$}}}
\def\nsl{\raise.15ex\hbox{$/$}\kern-.57em\hbox{$\nabla$}}
\def\sla{\raise.15ex\hbox{$/$}\kern-.57em\hbox{$\rightarrow$}}
\def\slla{\raise.15ex\hbox{$/$}\kern-.57em\hbox{$\lambda$}}
\def\gtwid{\raise.3ex\hbox{$>$\kern-.75em\lower1ex\hbox{$\sim$}}}
\def\ltwid{\raise.3ex\hbox{$<$\kern-.75em\lower1ex\hbox{$\sim$}}}

\def\12{{1\over2}}

\def\part{\partial}

\def\bethlogo{\vbox{\bf \line{\hrulefill}
    \kern-.5\baselineskip
    \line{\hrulefill\phantom{ ELIZABETH A. MASON }\hrulefill}
    \kern-.5\baselineskip
    \line{\hrulefill\hbox{ ELIZABETH A. MASON }\hrulefill}
    \kern-.5\baselineskip
    \line{\hrulefill\phantom{ 1411 Chino Street }\hrulefill}
    \kern-.5\baselineskip
    \line{\hrulefill\hbox{ 1411 Chino Street }\hrulefill}
    \kern-.5\baselineskip
    \line{\hrulefill\phantom{ Santa Barbara, CA 93101 }\hrulefill}
    \kern-.5\baselineskip
    \line{\hrulefill\hbox{ Santa Barbara, CA 93101 }\hrulefill}
    \kern-.5\baselineskip
    \line{\hrulefill\phantom{ (805) 962-2739 }\hrulefill}
    \kern-.5\baselineskip
    \line{\hrulefill\hbox{ (805) 962-2739 }\hrulefill}}}
\def\lisalogo{\vbox{\bf \line{\hrulefill}
    \kern-.5\baselineskip
    \line{\hrulefill\phantom{ LISA R. GOODFRIEND }\hrulefill}
    \kern-.5\baselineskip
    \line{\hrulefill\hbox{ LISA R. GOODFRIEND }\hrulefill}
    \kern-.5\baselineskip
    \line{\hrulefill\phantom{ 6646 Pasado }\hrulefill}
    \kern-.5\baselineskip
    \line{\hrulefill\hbox{ 6646 Pasado }\hrulefill}
    \kern-.5\baselineskip
    \line{\hrulefill\phantom{ Santa Barbara, CA 93108 }\hrulefill}
    \kern-.5\baselineskip
    \line{\hrulefill\hbox{ Santa Barbara, CA 93108 }\hrulefill}
    \kern-.5\baselineskip
    \line{\hrulefill\phantom{ (805) 962-2739 }\hrulefill}
    \kern-.5\baselineskip
    \line{\hrulefill\hbox{ (805) 962-2739 }\hrulefill}}}

\def\low#1{\lower.5ex\hbox{${}_#1$}}
\def\ltwid{\raise.3ex\hbox{$<$\kern-.75em\lower1ex\hbox{$\sim$}}}

\def\om{{\omega}}

\def\psl{\raise.15ex\hbox{$/$}\kern-.57em\hbox{$\partial$}}
\def\partt{\raise.15ex\hbox{$\widetilde$}{\kern-.37em\hbox{$\partial$}}}
\def\parts{\raise.15ex\hbox{$/$}{\kern-.6em\hbox{$\partial$}}}
\def\nablas{\raise.15ex\hbox{$/$}{\kern-.6em\hbox{$\nabla$}}}
\def\partw#1{\raise.15ex\hbox{$/$}{\kern-.6em\hbox{${#1}$}}}

\def\refto#1{$^{#1}$}

\def\si{{\sigma}}

\def\gtappr{{{\lower4pt\hbox{$>$} } \atop \widetilde{ \ \ \ }}}
\def\ltappr{{{\lower4pt\hbox{$<$} } \atop \widetilde{ \ \ \ }}}

\def\topppageno1{\global\footline={\hfil}\global\headline
={\ifnum\pageno<\firstpageno{\hfil}\else{\hss\twelverm --\ \folio
\ --\hss}\fi}}

\def\toppageno2{\global\footline={\hfil}\global\headline
={\ifnum\pageno<\firstpageno{\hfil}\else{\rightline{\hfill\hfill
\twelverm \ \folio
\ \hss}}\fi}}

\catcode`@=11
\newcount\tagnumber\tagnumber=0

\immediate\newwrite\eqnfile
\newif\if@qnfile\@qnfilefalse
\def\write@qn#1{}
\def\writenew@qn#1{}
\def\w@rnwrite#1{\write@qn{#1}\message{#1}}
\def\@rrwrite#1{\write@qn{#1}\errmessage{#1}}

\def\taghead#1{\gdef\t@ghead{#1}\global\tagnumber=0}
\def\t@ghead{}

\expandafter\def\csname @qnnum-3\endcsname
  {{\t@ghead\advance\tagnumber by -3\relax\number\tagnumber}}
\expandafter\def\csname @qnnum-2\endcsname
  {{\t@ghead\advance\tagnumber by -2\relax\number\tagnumber}}
\expandafter\def\csname @qnnum-1\endcsname
  {{\t@ghead\advance\tagnumber by -1\relax\number\tagnumber}}
\expandafter\def\csname @qnnum0\endcsname
  {\t@ghead\number\tagnumber}
\expandafter\def\csname @qnnum+1\endcsname
  {{\t@ghead\advance\tagnumber by 1\relax\number\tagnumber}}
\expandafter\def\csname @qnnum+2\endcsname
  {{\t@ghead\advance\tagnumber by 2\relax\number\tagnumber}}
\expandafter\def\csname @qnnum+3\endcsname
  {{\t@ghead\advance\tagnumber by 3\relax\number\tagnumber}}

\def\equationfile{%
  \@qnfiletrue\immediate\openout\eqnfile=\jobname.eqn%
  \def\write@qn##1{\if@qnfile\immediate\write\eqnfile{##1}\fi}
  \def\writenew@qn##1{\if@qnfile\immediate\write\eqnfile
    {\noexpand\tag{##1} = (\t@ghead\number\tagnumber)}\fi}
}

\def\callall#1{\xdef#1##1{#1{\noexpand\call{##1}}}}
\def\call#1{\each@rg\callr@nge{#1}}

\def\each@rg#1#2{{\let\thecsname=#1\expandafter\first@rg#2,\end,}}
\def\first@rg#1,{\thecsname{#1}\apply@rg}
\def\apply@rg#1,{\ifx\end#1\let\next=\relax%
\else,\thecsname{#1}\let\next=\apply@rg\fi\next}

\def\callr@nge#1{\calldor@nge#1-\end-}
\def\callr@ngeat#1\end-{#1}
\def\calldor@nge#1-#2-{\ifx\end#2\@qneatspace#1 %
  \else\calll@@p{#1}{#2}\callr@ngeat\fi}
\def\calll@@p#1#2{\ifnum#1>#2{\@rrwrite{Equation range #1-#2\space is bad.}
\errhelp{If you call a series of equations by the notation M-N, then M and
N must be integers, and N must be greater than or equal to M.}}\else%
 {\count0=#1\count1=#2\advance\count1
by1\relax\expandafter\@qncall\the\count0,%
  \loop\advance\count0 by1\relax%
    \ifnum\count0<\count1,\expandafter\@qncall\the\count0,%
  \repeat}\fi}

\def\@qneatspace#1#2 {\@qncall#1#2,}
\def\@qncall#1,{\ifunc@lled{#1}{\def\next{#1}\ifx\next\empty\else
  \w@rnwrite{Equation number \noexpand\(>>#1<<) has not been defined yet.}
  >>#1<<\fi}\else\csname @qnnum#1\endcsname\fi}

\let\eqnono=\eqno
\def\eqno(#1){\tag#1}
\def\tag#1$${\eqnono(\displayt@g#1 )$$}

\def\aligntag#1\endaligntag
  $${\gdef\tag##1\\{&(##1 )\cr}\eqalignno{#1\\}$$
  \gdef\tag##1$${\eqnono(\displayt@g##1 )$$}}

\def\eqalignno#1{\displ@y \tabskip\centering
  \halign to\displaywidth{\hfil$\displaystyle{##}$\tabskip\z@skip
    &$\displaystyle{{}##}$\hfil\tabskip\centering
    &\llap{$\displayt@gpar##$}\tabskip\z@skip\crcr
    #1\crcr}}

\def\displayt@gpar(#1){(\displayt@g#1 )}

\def\displayt@g#1 {\rm\ifunc@lled{#1}\global\advance\tagnumber by1
        {\def\next{#1}\ifx\next\empty\else\expandafter
        \xdef\csname @qnnum#1\endcsname{\t@ghead\number\tagnumber}\fi}%
  \writenew@qn{#1}\t@ghead\number\tagnumber\else
        {\edef\next{\t@ghead\number\tagnumber}%
        \expandafter\ifx\csname @qnnum#1\endcsname\next\else
        \w@rnwrite{Equation \noexpand\tag{#1} is a duplicate number.}\fi}%
  \csname @qnnum#1\endcsname\fi}

\def\ifunc@lled#1{\expandafter\ifx\csname @qnnum#1\endcsname\relax}

\let\@qnend=\end\gdef\end{\if@qnfile
\immediate\write16{Equation numbers written on []\jobname.EQN.}\fi\@qnend}

\catcode`@=12

\catcode`@=11
\newcount\r@fcount \r@fcount=0
\newcount\r@fcurr
\immediate\newwrite\reffile
\newif\ifr@ffile\r@ffilefalse
\def\w@rnwrite#1{\ifr@ffile\immediate\write\reffile{#1}\fi\message{#1}}

\def\writer@f#1>>{}
\def\referencefile{
  \r@ffiletrue\immediate\openout\reffile=\jobname.ref%
  \def\writer@f##1>>{\ifr@ffile\immediate\write\reffile%
    {\noexpand\refis{##1} = \csname r@fnum##1\endcsname = %
     \expandafter\expandafter\expandafter\strip@t\expandafter%
     \meaning\csname r@ftext\csname r@fnum##1\endcsname\endcsname}\fi}%
  \def\strip@t##1>>{}}

\def\citeall#1{\xdef#1##1{#1{\noexpand\cite{##1}}}}
\def\cite#1{\each@rg\citer@nge{#1}}

\def\each@rg#1#2{{\let\thecsname=#1\expandafter\first@rg#2,\end,}}
\def\first@rg#1,{\thecsname{#1}\apply@rg}	
\def\apply@rg#1,{\ifx\end#1\let\next=\relax
\else,\thecsname{#1}\let\next=\apply@rg\fi\next}

\def\citer@nge#1{\citedor@nge#1-\end-}	
\def\citer@ngeat#1\end-{#1}
\def\citedor@nge#1-#2-{\ifx\end#2\r@featspace#1 
  \else\citel@@p{#1}{#2}\citer@ngeat\fi}	
\def\citel@@p#1#2{\ifnum#1>#2{\errmessage{Reference range #1-#2\space is bad.}%
    \errhelp{If you cite a series of references by the notation M-N, then M and
    N must be integers, and N must be greater than or equal to M.}}\else%
 {\count0=#1\count1=#2\advance\count1
by1\relax\expandafter\r@fcite\the\count0,%
  \loop\advance\count0 by1\relax
    \ifnum\count0<\count1,\expandafter\r@fcite\the\count0,%
  \repeat}\fi}

\def\r@featspace#1#2 {\r@fcite#1#2,}	
\def\r@fcite#1,{\ifuncit@d{#1}
    \newr@f{#1}%
    \expandafter\gdef\csname r@ftext\number\r@fcount\endcsname%
                     {\message{Reference #1 to be supplied.}%
                      \writer@f#1>>#1 to be supplied.\par}%
 \fi%
 \csname r@fnum#1\endcsname}
\def\ifuncit@d#1{\expandafter\ifx\csname r@fnum#1\endcsname\relax}%
\def\newr@f#1{\global\advance\r@fcount by1%
    \expandafter\xdef\csname r@fnum#1\endcsname{\number\r@fcount}}

\let\r@fis=\refis			
\def\refis#1#2#3\par{\ifuncit@d{#1}
   \newr@f{#1}%
   \w@rnwrite{Reference #1=\number\r@fcount\space is not cited up to now.}\fi%
  \expandafter\gdef\csname r@ftext\csname r@fnum#1\endcsname\endcsname%
  {\writer@f#1>>#2#3\par}}

\def\ignoreuncited{
   \def\refis##1##2##3\par{\ifuncit@d{##1}%
    \else\expandafter\gdef\csname r@ftext\csname r@fnum##1\endcsname\endcsname%
     {\writer@f##1>>##2##3\par}\fi}}

\def\r@ferr{\endreferences\errmessage{I was expecting to see
\noexpand\endreferences before now;  I have inserted it here.}}
\let\r@ferences=\references
\def\references{\r@ferences\def\endmode{\r@ferr\par\endgroup}}

\let\endr@ferences=\endreferences
\def\endreferences{\r@fcurr=0
  {\loop\ifnum\r@fcurr<\r@fcount
    \advance\r@fcurr by1\relax\expandafter\r@fis\expandafter{\number\r@fcurr}%
    \csname r@ftext\number\r@fcurr\endcsname%
  \repeat}\gdef\r@ferr{}\endr@ferences}


\let\r@fend=\endpaper\gdef\endpaper{\ifr@ffile
\immediate\write16{Cross References written on []\jobname.REF.}\fi\r@fend}

\catcode`@=12

\def\reftorange#1#2#3{$^{\cite{#1}-\setbox0=\hbox{\cite{#2}}\cite{#3}}$}

\citeall\refto		
\citeall\ref		%
\citeall\Ref		%

\ignoreuncited
\tolerance=5000

\endtopmatter

\title Possible Realization of Odd Frequency Pairing in Heavy Fermion Compounds
\author P. Coleman$^1$, E. Miranda $^1$ and A. Tsvelik $^{2,3}$
\affil $^1$ Serin Physics Laboratory
Rutgers University
PO Box 849
Piscataway NJ 08855

\affil $^2$ Dept. of Physics
Jadwin Hall
Princeton University
Princeton NJ 08540

\abstract{ Using  Majorana Fermions to represent spins
we re-examine the Kondo Lattice model for heavy fermions.
The simplest decoupling procedure provides a
realization of odd frequency superconductivity, with
resonant pairing and surfaces of gap zeros.
Spin and charge coherence factors vanish linearly with the
energy on the Fermi surface, predicting a linear
specific heat, but a $T^3$ NMR relaxation rate.
Possible application
to heavy fermions is suggested.}

\vskip 1.0 truein PACS Nos. 75.20.Hr, 75.30.Mb, 75.40.Gb
\vskip 0.5 truein $^3$ Address after Sept 1, 1992 :
 Department of Physics,
Oxford University,
1 Keble Rd,
Oxford OX1 3NP,
UK.
\endtopmatter
\body

Though a decade and a half has passed since the discovery
of the heavy fermion metals and superconductors
\refto{ott,steglich},  many experimental anomalies
remain.  Whilst the basic theoretical picture of resonantly scattered
conduction electrons forming a highly renormalized f-band is not in
question, certain experimental features fit awkwardly into the
standard model\reftorange{nick}{rice, long, millis}{ auerbach}.  The
underlying nature of the interactions\refto{kuramoto}, the nature of
the pairing\refto{norman}, and the excitation spectrum of heavy
fermion insulators\reftorange{aeppli2}{ins12,ins22}{ins3} are three
areas of continuing uncertainty.

Conventional approaches to heavy fermion physics
represent the f-moments as fermions by  enforcing a ``Gutzwiller
constraint'' of unit occupancy
$n_f=1$ at each site.
This requires
a  projection of
the physical Hilbert space of the local moments from the larger
Hilbert space of pseudo-fermions: a task that is difficult to
do exactly, and usually only treated on the average.
In this letter we examine an approach to a simple Kondo lattice  model
for heavy fermions
that avoids these
difficulties.

Various new features are predicted  that
differ qualitatively from  the standard model of heavy fermion
behavior; most notably a
development of strong correlations between the spin and pair degrees
of freedom, forming a ground state where the conduction
electrons experience frequency dependent or ``resonant''
triplet pairing. The pairing fields  actually
diverge at low
frequencies as the inverse frequency, providing a first stable realization
of the phenomenon of
odd frequency pairing originally  considered by
Berezinskii\refto{berezinskii,abrahams}.
For a wide range of conditions, including the presence of spin-orbit
coupling, this theory predicts surfaces of gapless excitations, and
a linear specific heat that survives in the superconducting state.
Unlike a conventional superconductor, the charge and spin coherence
factors vanish on the pseudo-Fermi surface, giving rise to a
$T^3$ NMR relaxation. In this scenario, the
linear specific heat anomalies often
observed in heavy fermion
superconductors\refto{linear2} might be interpreted as intrinsic.

A key feature of our approach is the use of a special anticommuting
representation of spin one-half operators to describe the
magnetic excitations within the low-lying crystal field doublets of the heavy
fermion ions\refto{xtal}.
Recall that for individual
$S= 1/2$ objects, the Pauli matrices are anticommuting variables
$\{\si_a, \si_b\} = 2 \delta_{ab}$
and consequently can be treated as
real or Majorana ($\vec \si \dg = \vec \si$) Fermi fields.
Their Fermi statistics alone guarantee that
the spin operator
$
\vec S =- {i \over 4} \vec \si \times \vec \si
$
satisfies {\sl both} the SU(2) algebra $[S^a, S^b] = i \eps_{abc} S^c$
{\sl and}  the condition $\vec S^2 = 3/4$.
This feature can be generalized
to many sites, introducing a set of three component
anticommuting real vectors
$\vec \eta _i$ at each site i,
$$
\{\eta^a_i, \eta^b_j\} = \delta_{ij}\delta^{ab}\qquad\qquad(\eta_j^a =
{\eta\dg_j}^a)\qquad\qquad(a,b=1,2,3)
\eqno(comm)
$$
from which the spin operator at each site is constructed
$$
\vec S_j = -{i \over 2} \vec \eta_j \times \vec \eta_j\eqno(spin2)
$$
This ``Majorana'' representation of spin $1/2$ operators has a long
history\refto{history} in particle physics. Loosely speaking, the Majorana
fermions may be considered to be lattice generalizations of
{\sl anticommuting} Pauli operators $\vec \eta_j \equiv {1 \over \sqrt 2} \vec
\si_j$.
There is no constraint associated
with this representation,  for  the spin algebra {\sl and} the condition
$S=1/2$
are satisfied at each site, between all states of the Fock space\refto{weyl}.
In momentum space, the Bloch waves, $\vec \eta_{\vk} = \sum_j \eta_je^{-i \vk
\cdot\vec R_j}$ behave as conventional complex fermions, but
since $\eta\dg_{\vk} = \eta_{- \vk}$, the momentum lies
in one  half of the Brillouin zone. Finally note that since there
is no constraint, the trial ground-state  energy obtained from a
trial Hamiltonian is a {\it strict variational upper bound} on the true
ground-state energy.

Our basic model for a heavy fermion system is a  spin $1/2$
Kondo lattice model,
with a single band interacting with local f-moments
$\vec S_j$ in each unit cell.
Our simplified Hamiltonian is written
$$
H= H_{c} + \sum_j H_{int}[j]\eqno(kl)
$$
Here $H_c = \sum \eps_{\vk} \psi\dg_{\vk}\psi_{\vk}$ describes the conduction
band, and $\psi\dg_{\vk}= ( \psi\dg_{\vk \up} ,\psi\dg_{\vk \dw} ) $  is
a conduction electron spinor. The exchange interaction at each site $j$
is written in a tight binding
representation as
$$
H_{int}[j]= J (\psi\dg_{j\alpha} \vec \si_{\alpha\beta} \psi_{j \beta}) \cdot
\vec S_j \longrightarrow - {J \over 2}
\psi\dg_j[\vec \si_j \cdot\vec \eta_j]^2 \psi_j
$$
In a real heavy fermion system, we envisage that the
indices would refer to the conserved pseudospin indices of the
low lying magnetic manifold.
We have suppressed both the momentum
dependence and anisotropy of the coupling, using
$i\vec \si.(\vec \eta \times \vec \eta) =[\vec \eta \cdot
\vec \si]^2 - {3\over 2}$ to simplify the interaction.

We may now write the partition function as a path integral,
$
Z= \int_{\rm P}
e^{- \int_0^\beta {\cal L(\tau)}d\tau }
$
where
$$
{\cal L}(\tau) = \sum_{\vk }
\psi_{\vk} \dg\partial_{\tau} \psi_{\vk}+
\sum_{\vk \in {1 \over 2} BZ}\vec \eta_{\vk} \dg\partial_{\tau}\vec
\eta_{\vk}
+H_c + \sum_j H_{int}[j].
\eqno(lag2)
$$
Here we have
factorized  the interaction in terms of a fluctuating
two-component spinor $V\dg_j=(V^*_{\up}, V^*_{\dw})$
$$
H_{int}[j]=
\psi\dg_j ( \vec \si \cdot \vec \eta_j)V_j
+V\dg_j
( \vec \si \cdot \vec \eta_j)\psi_j + 2|V_j|^2/J
\eqno(int2)
$$
We are particularly interested in examining static mean field solutions
where
$$
V_j={V\over \sqrt{2}}\left( \matrix { z_{j\up} \cr z_{j\dw} \cr }
\right)\qquad\qquad
z_j\dg z_j=1
\eqno(static)
$$
To gain insight into this mean field theory,
let us integrate out the localized spin degrees of
freedom, represented by the Majorana fermions.
This introduces a
resonant self-energy into the electron propagators, containing
an isotropic component that builds the renormalized heavy fermion band
and an anisotropic  term,
compactly represented by the effective action
\def\iom{\omega_n}
$$
S_{c} =
\sum_{\{\vk, i\iom\}}\psi\dg_{\vk, \om}\biggl[ -\om+ \epsilon_{\vk}+
\Delta( \om)
\biggr]\psi_{\vk,\om}
+ S_{a}
\eqno(eff)
$$
where  $\Delta(\om ) = {V^2 \over 2  \om}$ determines the
strength of the resonant scattering.
The anisotropic term $S_a$
is written in a tight binding basis as
$$
S_{a} = - \sum_{ \{j, \ i \iom \}}{\Delta(\om)\over 2}
\biggl\{\psi\dg_{j,\om}[
1+\vec b_j \cdot \vec \si]\psi_{j,\om }+
\biggl[
\psi_{j,-\om} \bigl[i
\si_2\vec \si\cdot \vec d_j
\bigr]
\psi_{j,\om}+{\rm c.c.}
\biggr]
\biggr\}
\eqno(theworks)
$$
Here the
triad of orthogonal unit vectors
$\hat b =z\dg \vec \si z$, $\vec d=\hat x + i \hat y =
z^T[i \si_2 \vec \si] z$ define
the orientation of the order parameter.
The quantities
$$
\vec B_j(\om) = {\Delta(\om)\over 2}
 \hat b_j, \qquad\qquad
\vec \Delta_j(\om) ={\Delta({\om})\over 2} \hat d_j
\eqno(fields)
$$
may be interpreted as
resonant Weiss and triplet pairing fields, respectively.
Unlike earlier realizations of odd frequency triplet pairing\refto{berezinskii,
abrahams}, here
the odd frequency pairing field diverges at zero frequency, coupling
spin and triplet pair degrees
of freedom in one order parameter.

To simplify further discussion, we consider the
case of a bipartite lattice. Here, a stable mean field solution is obtained
with a {\sl staggered} order parameter, where for example
$\hat b$ is constant, and $\hat d = e^{i\vec Q \cdot \vec R_j}
\hat d_o$ is staggered commensurately with $\vec Q = ( \pi, \pi , \pi)$. In
this case the spinor $z_j= e^{i \vec Q \cdot \vec R_j/2}z_o$, where
$z_o=
\left(\matrix{1 \cr
0\cr}\right)$. Writing the conduction electron spinors
in terms of their four real components $\chi^{\lambda}(\vk)$
($\lambda=0,1,2,3$)
$$
\psi_j =  {1 \over \sqrt 2 }
\left\{
\chi^o_j+ i \vec \chi_j \cdot \vec \si
\right\}z_0\eqno(mft1)
$$
the mean field Hamiltonian takes the simple form
$$
H_{MF}
=\sum_{\vk\in {1 \over 2 } BZ}\left\{
\tilde \eps _{\vk} {\chi^{\lambda}_{\vk}}\dg\chi^{\lambda}_{\vk}
+ i V \biggl[ \vec \eta\dg_{\vk} \cdot \vec \chi_{\vk} - { \rm c.c.}\biggr]
+  \alpha_{\vk}n_{\vk}
\right\}
\eqno(mft2)
$$
where
$$
n_{\vk}=i\biggl[
({\chi^{3}_{\vk}} \dg
 \chi^o_{\vk} + {\chi^{2}_{\vk}}\dg \chi^1_{\vk}) - {\rm c.c}\biggr]
\eqno(charge)
$$
is the number operator of the state $ \vk $, written in the four component
basis and
$$
\tilde \eps_{\vk} = { 1 \over 2} ( \eps_{\vk + \vec Q /2 }
-\eps_{-\vk + \vec Q /2 }) ,
\qquad
\qquad
\alpha_{\vk} = { 1 \over 2} ( \eps_{\vk + \vec Q /2 }
+\eps_{-\vk + \vec Q /2 }) = - \mu
\eqno(oddeven)
$$
where the last equality holds only for a tight binding model.
Let us begin by considering the special case of half filling,
($\mu = 0$), for in this case the Hamiltonian is diagonal
in the  Majorana components $\lambda$, with excitation energies
$$\eqalign{
E_{\vk i} = &{ \tilde \eps_{\vk } \over 2}
\pm
\sqrt{
({ \tilde \eps_{\vk } \over 2} )^2 +  V^2
}\qquad\qquad(i=1,3)
\cr
E_{\vk 0} = & \tilde \eps_{\vk} \cr
}
\eqno(spec)
$$
corresponding to three hybridized gapful branches
and a forth gapless Majorana mode formed from a component of
the conduction band  that does not mix with the local moments.
With one
unpaired Majorana  fermion per unit cell, the corresponding
Fermi surface $\tilde \eps_{\vk}=0$
spans precisely one half of the Brillouin zone:
$V_{FS}/(2 \pi)^3 = {1 \over 2}$.
This  counting argument guarantees that the gapless Fermi surface
persists in the presence of particle hole asymmetry ($\mu\neq 0$)
or a spin dependent kinetic energy associated with
spin-orbit coupling.

For our particular choice of $z_o$, the up electrons are ``paired'', whilst
the ``down'' electrons are unpaired with a gapped excitation
spectrum (Fig. 1.). In
a Nambu notation,
their
propagators are
\def\e{\tilde \epsilon_{\vk}}
$$
G_{ \si}(\om,\vk) =\left\{\eqalign{
 [(\omega - \ul{\eps}_{\vk} - \Delta(\om)(1 +
\ul{\tau}_1)]^{-1}\qquad\qquad&(\si=\up)\cr
 [(\omega - \ul{\eps}_{\vk} - 2\Delta(\om)]^{-1}\qquad\qquad &(\si=\dw)\cr
}\right.\eqno(bluch)
$$
where $\ul{\eps}_{\vk} = \e - \mu \ul{\tau}_3$. The
density of states for the ``up'' electrons is
$$
\rho_{\up}(\om) =
\left\{\eqalign{
&{ \rho \over 2} \biggl(
1 + { \mu  \omega \over
[V^4 /4 + \mu^2 \omega^2 ]^{1 \over 2}
}\biggr)\quad (\vert \omega\vert < T_{K})\cr
&\rho\qquad\quad\qquad\qquad \qquad\qquad(\vert \omega\vert  > T_{K})\cr
}\right. ,
\eqno(density)
$$
where $T_K= {V^2 \over D[1-(\mu/D)^2]}$ is the indirect gap associated with the
excitation spectrum \(spec).

Unlike conventional pairing, the charge and spin coherence factors
of these ``Majorana'' quasiparticles are strongly energy dependent.
Near the Fermi surface,
the gapless quasiparticle operators can be written as
$$
a_{\vk\up}\dg = Z^{1 \over 2}\bigl[u_{\vk} \psi\dg_{\vk \up} + v_{\vk}
\psi_{-\vk \up} \bigr]+ \bigl(1-Z\bigr)^{1 \over 2}\eta\dg_{3\vk}
\eqno(qp)
$$
where $Z^{-1}=1+{\mu^2 \over V^2}$ is a quasiparticle renormalization
constant and the Bogoliubov coefficients are determined by $u^2 + v^2 =1$,
$$
u_{\vk}^2 = {1 \over 2}\biggl[1 + { \mu \ {\rm sgn}(E_{\vk})\over \sqrt{
\Delta(E_{\vk})^2 + \mu^2 }}\biggr]
\eqno(cohfact)
$$
Spin and charge coherence factors are then given by
$$
\left.\eqalign{
&\langle \vk^- \vert
\rho_{\vq}\vert \vk^+ \rangle\cr
& \langle\vk^- \vert  \si^z_{\vq}\vert \vk^+ \rangle\cr}\right\}
=
Z\bigl[u_{\vk^-}u_{\vk^+} - v_{\vk^-} v_{\vk^+}\bigr]=
(E_{\vk^+}+E_{\vk^-})
{Z\mu \over V^2}
\eqno(charge1)
$$
These quasiparticles thus form a pseudogap
where spin and charge matrix elements
{\sl vanish} on the Fermi surface and
grow {\sl linearly} with energy,
In the special particle-hole
symmetric case ($\mu=0$), these coherence factors vanish {\it throughout}
the gap, forming  a neutral band of excitations
that  only conduct heat.
Since the paramagnetic spin and charge
response functions of the quasiparticle fluid are
proportional to the square of these matrix
elements, the corresponding local response functions
to grow quadratically with energy
$$
{{\chi''_{sp, ch }(\omega)\over \omega}} \propto \biggl({\omega\over T_K}
\biggr)^2 \biggl({\mu \over D}\biggr)^2
\eqno(resp)
$$
This unusual energy dependence of matrix elements
permits this state to mimic one with
constant coherence factors, but a {\sl linear}
density of states (line of gap zeros).

We briefly list the main consequences of these results:

\item{i)} A large
quasiparticle thermal conductivity in the absence of a quasiparticle
contribution to the thermopower
and electrical conductivity.

\item{ii)} Linear specific heat coefficient of magnitude
$\gamma={1 \over 4}\gamma_n
(1 + \mu^2/V^2)$ where $\gamma_n $ is the linear specific heat in the absence
of the local moments.
As $\mu$ varies   $\gamma$ can
vary between values characteristic of a conventional metal, and
values characteristic of a heavy fermion metal.

\item{iii)}A $T^3$  component to
the NMR relaxation rate  superimposed upon
an activated background.
$$
{1 \over T_1T}\propto \left(1 + {V^2\over T_K^2} \right)e^{-( T_K /T)}
+ { \pi^2 \over 6} \left({T^2 \mu^2 \over V^4}\right)\eqno(nmr)
$$
Since the spin matrix elements $\langle \eps\vert S^{\pm}\vert\eps\ra =0$,
the $T^3$  response is anisotropic
and vanishes  when
the applied field is parallel to the $\hat b$ axis.

Finally, we should like to mention the collective properties of this
state. Past studies of odd-frequency
pairing have encountered a negative phase stiffness \refto{private}. In our
mean field theory, the phase has
``coiled up'' into a staggered configuration: this stabilizes
the state and develops a positive phase stiffness. To compute the London
response to a vector potential $\vec A$, we replace $\eps_{\vk}
\longrightarrow \eps_{\vk - e \vec A \tau_3}$.
The London Kernel $\Lambda \delta_{ab} = \nabla^2_{A_aA_b}F[\vec A]$ is then
$$
\Lambda =  - { e^ 2 v_F^2 T \over 6} \sum_{\kappa}
{\rm Tr} \biggl[
{\cal G}_{c \up}(\kappa)^2-
{\cal G}_{c \up}(\kappa)\tau_3{\cal G}_{c \up}(\kappa)\tau_3
\biggr]
\eqno(stiff)
$$
where the minus sign is a result of the staggered phase.
Carrying out the energy integral, at $T=0$
$$
\Lambda
= {Ne^2 \over 4 m}  \int_o^{T_K} {d \omega}
{ \Delta({\om})^2 \over ( \Delta({\om})^2 + \mu^2 )^{3 \over 2}}
\eqno(stiff5)
$$
where we have set ${N \over 2 m} \equiv
{\rho v_F^2 \over 3}$.
In the special case of $\mu=0$, this integral simplifies
to
$
\Lambda	 =
{Ne^2 \over  m} \left( {
T_K\over 4 D }\right)
$
This small phase stiffness is consistent with the large coherence
lengths $\lambda_L^{-2}= \mu_o \Lambda$
of heavy fermion superconductors, and may be regarded
as a consequence of a  condensation of ``heavy electrons'' with mass
$m^*= m{D\over T_K}$.

Macroscopic properties of the paired state are governed by slow rotations
of the order parameter.
In the absence of anisotropy, the
long-wavelength action is fully spin-rotationally invariant, given by a
$U(1)_{charge}\times SU(2)_{spin}$ sigma model.
Despite the phase stiffness,
it
can not support a supercurrent without anisotropy,
for the vacuum is  not
topologically against the creation of vortices:
uniform gradients
of the phase can continuously develop to cancel an externally
applied vector potential by
twisting the order parameter
``into the third dimension''\refto{mermin}.
Anisotropy aligns the order parameter with the crystal
axes, lowering  the symmetry to a
$U(1)_{charge}$ $X-Y$  model, where a supercurrent is topologically
stable.
In the special case of half filling the order parameter
can be uniformly rotated in particle-hole space
$
\left( \matrix{z_{\up}\cr z^*_{\dw} \cr}\right)
\rarrow g \left( \matrix{z_{\up}\cr z^*_{\dw} \cr}\right) $
where $g$ is an $SU(2)$ matrix. Now the long-wavelength action
possess an $SU(2)_{charge}$ symmetry,
and again, does not support a persistent current.
Since the half filled state has a gap to both spin and charge
excitations, this suggests this state will be
insulating: a ``superconductor''
with a vanishing critical current and upper-critical field.

Several questions of a technical nature surround our
simple mean field theory.
One  performance benchmark of our mft is provided by
the one impurity Kondo model: here we may compare its performance
with the exact results and the well established large $N$ mean field
theory (mft)\refto{nick}.
For this model,
the Majorana mft correctly yields  a unitary
${\pi\over 2}$ phase shift for the scattered electrons; it also
predicts an enhanced isotropic
susceptibility and linear specific heat: features
consistent with the Fermi liquid
fixed point. The mean field
Wilson ratio ${\chi\over \gamma}
=8/3$ compares more favorably with the
exact value ${\chi\over \gamma}=2$ than the large $N$ mft, where
$\left({\chi\over \gamma}\right)_{N\rarrow\infty}=1$.
As in the
large $N$ approach, RPA
fluctuations in the phase variables develop
power-law correlations the mean field order parameter,
but
here there is no local continuous gauge symmetry so
power-law correlations  are physically
manifested
as  long-time correlations of the  spin-charge
operator $\vec M(t) = \vec S(t) \rho(t)$.
Verification of such correlations in the one impurity
model would provide an independent test of  our technique.
Beyond the one impurity model, it remains to be seen whether our
approach can also recover the normal phase by a careful treatment of
these fluctuations.

Experimentally, the strong frequency dependence of
coherence factors in our theory
may help reconcile the observation of large linear specific heats
and thermal conductivities in heavy fermion
superconductors with the consistent {\sl absence} of  a corresponding Korringa
NMR relaxation normally associated with gapless superconductivity.
There is  also a possible
link with
Kondo insulators
\refto{ins12, ins22}.
In $CeNiSn$, a $T^3$ NMR relaxation rate and
pseudogap have been observed\refto{ins22}, features consistent with
incipient odd pairing.
It would be interesting to
measure and compare the
thermal conductivity of $CeNiSn$ with $LaNiSn$, to check for anisotropies in
the $T^3$ NMR response and test for a possible
proximity effect with other heavy fermion superconductors.

In conclusion, we have examined an alternative treatment of the Kondo
lattice model for heavy fermions that uses a Majorana representation
of the spins. Our theory predicts a low temperature ground-state
with odd frequency triplet pairing and surfaces of gapless
neutral excitations. Spin and charge coherence factors
vanish on the Fermi surface, predicting
an intrinsic
thermal conductivity and linear specific heat that coexist-exist with
a $T^3$ NMR relaxation rate.
Independently of these issues, it provides a first
stable realization of Berezinskii's odd frequency pairing.

We would particularly like to thank E. Abrahams
and
P. W. Anderson for discussions related to this work.
Discussions with N. Andrei,
A. V. Balatsky, D. Khmelnitskii, G. Kotliar,
G. Lonzarich  and A. Ramirez are also gratefully acknowledged.
Part of the work was supported by NSF grants DMR-89-13692 and  NSF 2456276.
P. C. is a Sloan Foundation Fellow. E. M. was supported by a grant
from CNPq, Brazil.

\references

\refis{ott}K. Andres   , J. Graebner \& H. R. Ott., \prl 35, 1779,
1975.

\refis{steglich}F. Steglich , J. Aarts. C. D. Bredl, W. Leike,
D. E. Meshida, W. Franz \& H. Sch\"afer, \prl 43, 1892, 1976.

\refis{nick}N. Read \&  D. M. Newns, \jpc 29, L1055, 1983 ;
N.Read, \jpc 18, 2051, 1985.

\refis{rice} T. M. Rice and K. Ueda, \prb 34, 6420, 1986; C. M. Varma,
W. Weber and L. J. Randall, \prb 33, 1015, 1986.

\refis{long}P. Coleman, \prb  35, 5072, 1987.

\refis{millis}
A.J. Millis and P.A. Lee, \prb 35, 3394, 1986.

\refis{auerbach}
A. Auerbach and K.Levin,\prl 57, 877, 1986.

\refis{kuramoto}Y. Kuramoto and K. Miyake,
\journal Prog. Theo. Phys. Suppl., 108, 199, 1992.

\refis{norman}M. R. Norman, \journal Physica, C194, 203, 1992.

\refis{aeppli2}T. Takabatake et al, \prb, 45, 5740, 1992;
T. Mason et al, \prl 69, 490, 1992.

\refis{ins1}M. F, Hundley, P. C. Canfield, J. D. Thompson, Z. Fisk
and J. M. Lawrence, \prb 42, 6842, 1990.

\refis{ins2}F. G. Aliev, V. V. Moschalkov,
V. V. Kozyrkov, M. K. Zalyalyutdinov,
V. V. Pryadum and R. V. Scolozdra, \jmmm 76-77, 295, 1988.

\refis{ins3}S. Doniach and P. Fazekas, {\sl Phil. Mag. }, {\bf 65B}
1171 (1992).

\refis{berezinskii}V. L. Berezinskii, \journal JETP Lett. , 20, 287,
1974.

\refis{abrahams}For recent interest in odd-frequency
pairing, see
E. Abrahams, A. V. Balatsky,
\prb 45, 13125, 1992;
F. Mila and E. Abrahams, \prl 67, 2379, 1991.

\refis{linear}U. Rauschwalbe, U. Ahlheim, C. D. Bredl, H. M. Meyer and
F. Steglich \jmmm 63\&64, 447, 1987; R. A. Fisher, S. Kim, B. F. Woodfield,
N. E. Phillips, L. Taillefer, K. Hasselbach, J. Floquet, A. L. Giorgi and
J. L. Smith, \prl 62, 1411, 1989.

\refis{xtal}Many heavy fermion compounds clearly show Schottky
anomalies in their specific heat where the entropy integral
beneath corresponds to the suppression of magnetic fluctuations
into the higher crystal field states. See for example,
F. Rietschel et al,
\jmmm 76\&77, 105, 1988, R. Felten et al, \journal Eur. Phys. Let.,
2 , 323, 1986.

\refis{history}J. L. Martin, \journal Proc. Roy. Soc., A 251, 536, 1959;
R. Casalbuoni, \journal Nuovo Cimento, 33A, 389, 1976;
F. A. Berezin \& M. S. Marinov, \journal Ann. Phys., 104, 336, 1977.

\refis{weyl}See e.g.
R. Brauer and H. Weyl, \journal Amer. J. Math, 57, 425, 1935.
The Majorana  Fock space is simply constructed in momentum space.
Though this space is larger than
the conventional Hilbert space of spin $1/2$ operators that
commute at different sites, it remains faithful
to the algebra by {\sl replicating} the conventional Hilbert space
$2^{n}$ times for an even number of $2n$ sites.

\refis{private}E. Abrahams (private communication).

\refis{mermin}N.D. Mermin, \journal Rev. Mod. Phys., 51, 591, 1979.

\refis{ins12}M. F. Hundley et al, \prb 42, 6842, 1990.

\refis{ins22}F. G. Aliev et al, \jmmm 76-77, 295, 1988.

\refis{aeppli}T. Takabatake, M. Nagasawa, H. Fujii, G. Kido,
M. Nohara, S. Nishigori, T. Suzuki, T. Fujita, R, Helfrich, U. Ahlheim,
K. Fraas, C. Geibel, and F. Steglich, \prb, 45, 5740, 1992;
T. Mason, G. Aeppli, A. P. Ramirez, K. N. Clausen, C. Broholm,
N. Stucheli, E. Bucher, T. T. M. Palstra, \prl 69, 490, 1992.

\refis{linear2}U. Rauschwalbe et al,\jmmm 63\&64, 447, 1987; R. A. Fisher et
al,  \prl 62, 1411, 1989.

\endreferences

\figurecaptions

\noindent {\bf Fig. 1. } Excitation spectrum of mean field theory
for $\mu=0$, showing three-fold degenerate gapped excitations
and a gapless Majorana band. Inset, conduction electron density of
states for up ($\up$) and down ($\dw$) electrons.

\endfigurecaptions

\endit
\end